\documentclass[prd, amsfonts, onecolumn, nofootinbib, showpacs]{revtex4}
\usepackage{graphicx, epsfig}
\usepackage{color}
\usepackage{amsmath}
\usepackage{amssymb}
\newcommand{\be}{\begin{equation}}
\newcommand{\ee}{\end{equation}}
\newcommand{\bea}{\begin{eqnarray}}
\newcommand{\eea}{\end{eqnarray}}

\newcommand{\gapp}{\mathrel{\raise.3ex\hbox{$>$}\mkern-14mu
\lower0.6ex\hbox{$\sim$}}}
\newcommand{\lapp}{\mathrel{\raise.3ex\hbox{$<$}\mkern-14mu
\lower0.6ex\hbox{$\sim$}}}
\def\bbox{{\,\lower0.9pt\vbox{\hrule \hbox{\vrule height 0.2 cm
\hskip 0.2 cm \vrule  height 0.2 cm}\hrule}\,}}

\begin{document}
\title{Primordial scalar gravitational waves produced at the QCD phase transition due to the trace anomaly}
\author{De-Chang Dai$^{1,2}$\footnote{corresponding authors: D. Stojkovic, D. Dai,\\ email:   ds77@buffalo.edu  diedachung@gmail.com      \label{fnlabel}}, Dejan Stojkovic$^3$}
\affiliation{$^1$ Center for Gravity and Cosmology, School of Physics Science and Technology, Yangzhou University, 180 Siwangting Road, Yangzhou City, Jiangsu Province, P.R. China 225002}
\affiliation{ $^2$ CERCA/Department of Physics/ISO, Case Western Reserve University, Cleveland OH 44106-7079}
\affiliation{ $^3$ HEPCOS, Department of Physics, SUNY at Buffalo, Buffalo, NY 14260-1500, US}

\begin{abstract}
\widetext
Relying only on the standard model of elementary particles and gravity, we study the details of a new source of gravitational waves whose origin is in quantum physics. Namely, it is well known that massless fields in curved backgrounds suffer from the so-called ``trace anomaly". This anomaly can be cast in terms of new scalar degrees of freedom which take account of macroscopic effects of quantum matter in gravitational fields.  The linearized effective action for these fields describes scalar (as opposed to transverse) gravitational waves, which are absent in Einstein's theory. Since these new degrees of freedom couple directly to the gauge field scalars in QCD, the epoch of the QCD phase transition in early universe is a possible source of primordial cosmological gravitational radiation. While the anomaly is most likely fully unsuppressed at the QCD densities (temperature is much higher than the u and d quark masses), just to be careful we introduced the window function which cuts-off very low frequencies where the anomaly effect might be suppressed. We then calculated the characteristic strain of the  properly adjusted gravitational waves signal today. The region of the parameter space with no window function gives a stronger signal, and both the strain and the frequencies fall within the sensitivity of the near future gravitational wave experiments (e.g. LISA and The Big Bang Observer). The possibility that one can study quantum physics with gravitational wave astronomy even in principle is exciting, and will be of value for future endeavors in this field.
\end{abstract}


\pacs{}
\maketitle

\section{Introduction}
Recent detection of gravitational waves opened a new window for exploration of our universe \cite{Abbott:2016blz}. For the first time we can directly study violent events like black holes (and other compact objects) mergers \cite{Mandic:2016lcn,Bhagwat:2016ntk,Clesse:2016ajp}, or collapse of massive stars \cite{Crocker:2017agi}. What is perhaps even more important, primordial gravitational waves can give us information about the early universe that is impossible to obtain from photons even in theory. For example, gravitons emitted during Hawking evaporation of primordial black holes should be observed as (appropriately redshifted) gravitational waves today \cite{Dong:2015yjs,Anantua:2008am}. This is perhaps our best bet to ever observe effects of Hawking radiation from astrophysical black holes. We can also learn about the high energy fundamental physics above the electroweak phase transition. Namely, if the dimensionality of the space-time changes at high temperatures, then the physics of the propagation of gravitational waves might change. In the context of the so-called ``vanishing dimensions" models, the solution to the standard model hierarchy problem requires the reduction of number of dimensions just above the electroweak scale\cite{Anchordoqui:2010er}. Since there are no propagating degrees of freedom in Einstein's gravity in less than three spatial dimensions, that would imply a cut-off at some frequency in the spectrum of primordial gravitational waves \cite{Mureika:2011bv,Stojkovic:2014lha}. Alternative theories of gravity have been analyzed in \cite{Yunes:2013dva}. For other applications, see review in \cite{Yunes:2016jcc}.

The goal of this paper is to study some unique predictions of the standard model of elementary particles coupled to gravity.
It is well known that massless fields in curved backgrounds suffer from the so-called ``trace anomaly". This anomaly induces the non-local effective action which however can be cast into local form with the help of some additional scalar degrees of freedom \cite{Mottola:2006ew,Giannotti:2008cv}. These fields take account of macroscopic effects of quantum matter in gravitational fields, which are not contained in the local metric description of Einstein's theory. Despite the fact that the existence of these fields follow straight from the standard model and general relativity (with no exotic physics), their consequences and phenomenology have not been extensively studied so far.
The linearized effective action for these fields describes scalar gravitational waves, which are absent in Einstein's theory.
Since they couple directly to the gauge field scalars, such as $G_{\mu \nu}^a G^{a \mu \nu}$ in the quantum chromodynamics (QCD), mergers of dense sources like neutron stars can give rise to these scalar gravitational waves.  Some rough estimates for dense sources were given in \cite{Mottola:2016mpl}. In this paper we study an alternative source of the scalar gravitational waves. Namely, in early universe at temperatures higher than $150$MeV, the QCD anomaly becomes unsuppressed, at least in some frequency range. This epoch of the QCD phase transition is a possible source of  primordial cosmological (scalar) gravitational radiation, in addition to the standard tensor gravitational waves \cite{Witten:1984rs}. .

\section{Homogeneous QCD phase transition}

We first give a brief overview of the the QCD phase transition with the relevant numbers, which will  be relevant for calculating the characteristics of the gravitational waves signal.
At temperatures above the QCD phase transition temperature ($T_c \approx 150 MeV$), the universe is full of free quarks, gluons and photons. At these temperatures the first two generations of quarks (u and d) are highly relativistic, and can be treated as massless since the temperature of the environment is much higher than their masses. The Hubble time at the QCD phase transition ($t_{QCD}\approx 10^{-5}s$) is much longer than the relaxation time scale for particle interactions, so the these particles are in thermal and chemical equilibrium. As the temperature of the universe decreases, some quarks and gluons condense to create hadronic matter. It takes about $0.1 \, t_{QCD} \approx 10^{-6}s$ for this phase transition to be completed.

\begin{figure}
   \centering
\includegraphics[width=8cm]{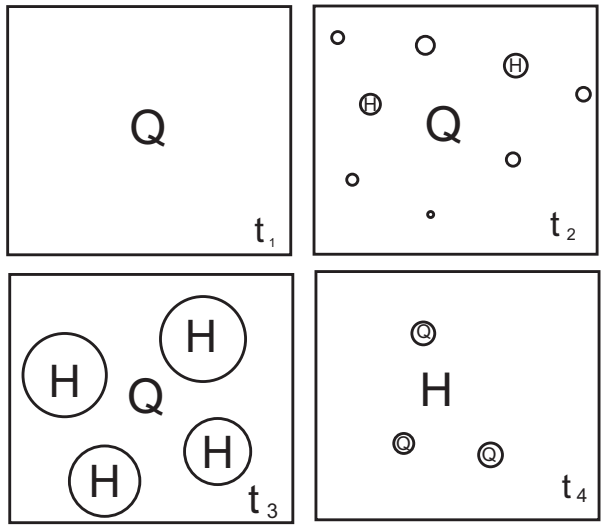}
\caption{
Early universe is dominated by radiation. Before the QCD phase transition, the universe is full of free quark-gluon matter (labeled by $Q$ in the figure), while hadrons are absent. During the first order phase transition, at some early time, $t_2$, some hadronic bubbles (labeled by $H$ in the figure) appear after a brief period of supercooling. These bubbles appear suddenly and release their latent heat to reheat the space outside of the bubbles. These small bubbles cover about $1\%$ of volume of the universe and then quench. The average distance between the bubbles is $d_{nuc}$. After that, they grow following adiabatic expansion of the universe. At time  $t_3$, the bubbles grow to a radius of about $d_{nuc}$. At that time, hadronic bubbles (H) occupy most of the space.  At $t_4$, the hadronic bubbles merge, and only very few free quark droplets are found in the hot spots.
}
\label{phase}
\end{figure}

If the QCD phase transition is a first order transition, it proceeds via bubble nucleation \cite{Hogan:1984hx,DeGrand:1984uq,Boyanovsky:2006bf}. If there is no impure matter in the universe to create an early nucleation core, the QCD phase transition will not happen immediately when the temperature drops to $T=T_{QCD}$. Instead, the hadronic bubbles nucleate after a short period of supercooling, $t_{sc}\approx 10^{-3}t_{QCD}$. Once small hadronic bubbles are formed, their bubble walls expand by weak deflagration \cite{DeGrand:1984uq,Ignatius:1993qn,Ignatius:1994fr,KurkiSuonio:1995pp,Kajantie:1986hq}. The deflagration fronts move at the speed $v_{def}$. The bubble volume grows very quickly with time
\begin{equation}
\label{Vb}
V_{bubble}=\frac{4\pi}{3} \Big( v_{def}\Delta t\Big)^3 ,
\end{equation}
where $\Delta t$ is time elapsed since the bubble formation. The period of bubble deflagrating growth is finished after $\Delta t_{nuc}\approx 10^{-6}t_{QCD}$. The phase transition releases latent heat and reheats the nearby region. The heat is transferred with the speed $v_{heat}$. The latent heat prevents any additional nucleation in these regions. Therefore, the average distance between the bubbles is $d_{nuc}\approx 2v_{heat}\Delta t_{nuc}\approx 1\text{cm}$ (this period is labeled as $t_2$ in Fig.~\ref{phase}). However, $d_{nuc}$ is about $1$m in \cite{Kajantie:1986hq}, so we will use both values to explore the whole parameter space. The bubble radius is about $R_{bubble}\approx v_{def} \Delta t_{nuc}$. The supercooled regions cover about $1\%$ of the volume of the universe, so their volume is about $10^{-2}\frac{4\pi}{3}(\frac{d_{nuc}}{2})^3$.

The bubble growth rate after deflagration slows down and is dominated by the universe expansion until the bubble grows to the size of $d_{nuc}/2$ (this period is labeled as $t_3$ in Fig.~\ref{phase}). At time $t_4$, the bubbles merge and leave very few free quark-gluon drops. After the deflagration phase, the hadron bubble grows because the universe is cooling down. If the supercooling is neglected, the volume fraction of matter in the hadron phase can be written as \cite{Kajantie:1986hq}
\begin{equation}
f(t)=1-\frac{1}{4(r-1)}\Bigg(\tan^2\Big(\arctan\sqrt{4r-1}+\frac{3\chi(t_i-t)}{2\sqrt{r-1}}\Big)-3\Bigg) ,
\end{equation}
based on the bag model. $t_i$ is the initial time when the QCD phase transition started, $\chi=\sqrt{8\pi G B}=\frac{1}{36\mu \text{sec}}(\frac{T_c}{200MeV})^2$, and $r$ is set to be $3$ in \cite{Kajantie:1986hq}. Here, $B$ is the bag energy. In this period, the single bubble's volume increases with time
\begin{equation}
\label{Vb1}
V_{bubble}=V_0 f(t)
\end{equation}
where, $V_0\approx \frac{4\pi}{3}(\frac{d_{nuc}}{2})^3$. In this formula, we neglect the contribution to the volume from the deflagration period because it is much smaller.

This is a basic picture of the QCD phase transition. One may also consider temperature fluctuations which can cause inhomogeneous nucleation \cite{Ignatius:2000gv}. However, this will not change the formation process of the hadronic bubbles.

\section{Scalar gravitational waves from the trace anomaly}
\label{sgv}

Any cosmological first order phase transition can produce gravitational waves in three different ways --- through the bubble collisions \cite{Kosowsky:1992rz,Kosowsky:1992vn,Kamionkowski:1993fg}, production of sound waves \cite{Hindmarsh:2013xza}, and magnetohydrodynamic turbulence \cite{Kosowsky:2001xp}. These production mechanisms have been previously studied in \cite{Caprini:2007xq,Huber:2008hg,Jinno:2015doa}. In some cases, gravitational waves are strong enough to be detected by the future gravitational wave detectors \cite{Ahmadvand:2017xrw,Aoki:2017aws}. In addition, the International Pulsar Timing Array can detect the gravitational wave generated by QCD bubble collisions\cite{Caprini:2010xv}.  Apart from gravitational wave created by the isotropic mass distribution, the QCD phase transition can also change the primordial gravitational waves power spectrum \cite{Schwarz:1997gv,Schettler2011}. If primordial gravitational waves are detected, they could provide an evidence for inflation and/or phase transitions in the early universe.

So far, tensor mode (or transverse) gravitational waves have been very well studied in the literature, unlike the scalar mode gravitational waves. One of the reasons is that it is not easy to generate scalar mode gravitational waves. Two possible sources are high energy/density QCD or QED states which at the quantum level suffer from the so-called ``trace anomaly". These can be naturally achieved in cores of dense (neutron) stars or in the very early universe. Here we study the possibility of the scalar mode gravitational wave production during the QCD phase transition in the early universe.

It is well known that quantum massless fields propagating in classical curved backgrounds suffer from the ``gravitational trace anomaly".
Simply, the trace of the stress energy tensor for the massless field, which vanishes in Minkowski space, acquires additional terms due to the curvature of the background and it does not vanish.
The general form of this gravitational trace anomaly in four space-time dimensions, is given by \cite{Mottola:2016mpl}
\begin{equation} \label{tmunu}
T^\mu_\mu = bC^2 +b' (E-\frac{2}{3} \Box R)+b'' \Box R +\sum_i \beta_i L_i ,
\end{equation}
where,
\begin{eqnarray}
&E=R_{\alpha\beta\gamma\delta}R^{\alpha\beta\gamma\delta}-R_{\alpha\beta}R^{\alpha\beta}+R^2\\
&C^2=R_{\alpha\beta\gamma\delta}R^{\alpha\beta\gamma\delta}-2R_{\alpha\beta}R^{\alpha\beta}+\frac{1}{3}R^2 .
\end{eqnarray}
Here, $L_i$ is the Lagrangian of a massless gauge field, while $E$ and $C$ are given in terms of curvature invariants. In the context of the standard model that we are concerned about here, $L_i$  is either the quantum electrodynamics (QED) or quantum chromodynamics (QCD) Lagrangian. Parameters $b$, $b'$, $b''$ and $\beta_i$ are some dimensionless constants. In particular,
\begin{eqnarray}
&b=\frac{\hbar}{120(4\pi)^2}(N_s+6N_f+12N_v)\\
&b'=-\frac{\hbar}{360(4\pi)^2}(N_s+11N_f+62N_v) ,
\end{eqnarray}
where $N_s$, $N_f$ and $N_v$ represent the number of free conformal scalars, four-component Dirac fermions, and vectors respectively.
The coefficients $b$ and $b'$ cannot be removed by any local counterterms and represent a true anomaly. The coefficient  $b''$ can be adjusted or set to zero.
The coefficients  $\beta_i$ are the $\beta$-functions of the corresponding gauge couplings in the Lagrangians $L_i$.

The anomalous terms on the right hand side of Eq.~(\ref{tmunu}) can be described by a non-local effective action. However the non-local action can be cast into local form with the help of an additional scalar degree of freedom, $\phi$. This field take account of macroscopic effects of quantum matter in gravitational fields, which are not contained in the local metric description of Einstein's theory.
The complete local semi-classical effective action for the gravity plus the anomaly is \cite{Mottola:2016mpl}
\begin{equation}
S_{eff}=S_{EH}(g)+S_{anom}(g,\phi) ,
\end{equation}
where $S_{EH}(g)$ is the Einstein-Hilbert term
\begin{equation} \label{eha}
S_{EH}(g)=\frac{1}{16G}\int d^4 x \sqrt{-g}(R-2\Lambda) .
\end{equation}
Here, the speed of light is taken to be $c=1$. $S_{anom}(g,\phi)$ is a local effective action
\begin{eqnarray}
&S_{anom}(g,\phi)=-\frac{b'}{2}\int d^4 x \sqrt{-g}\Big[(\Box \phi)^2-2(R^{\mu\nu}-\frac{1}{3}Rg^{\mu\nu})\triangledown_\mu\phi\triangledown_\nu\phi\Big]\\
&+\frac{1}{2}\int d^4 x \sqrt{-g}\Big[b'(E-\frac{2}{3}\Box R)+bC^2+\sum_i\beta_i L_i\Big]\phi
\end{eqnarray}

 In general, one should add the contribution from the scalar field to the total energy density of the universe. The energy momentum tensor for the auxiliary scalar field in early universe is (see e.g. \cite{Anderson:2009ci})
\begin{equation}
T^{anom}_{\alpha\beta}=6b' H^4 g_{\alpha\beta} ,
\end{equation}
where $H$ is the early time Hubble parameter. This form is similar to the energy momentum in de Sitter spacetime. During the QCD phase transition period, $H\approx 1/t_{QCD}$, the energy density is
\begin{equation}
\rho^{anom}= -6b' H^4 \sim  10^{-68}(MeV)^4 .
\end{equation}
 This values is far below the energy density of the ordinary radiation, $\sim T_c^4$, so we can safely neglect the scalar field's thermal energy density. In addition, it was argued  that coupling to the extra scalar field may cause infrared divergencies due to state dependent variations on the horizon scale \cite{Anderson:2009ci}. However, the effect that we study here is well inside the causal distance, so this divergence at the horizon scales may be neglected too. Finally, during the QCD phase transition, the universe is radiation dominated, so the cosmological constant (dark energy) effects can be neglected at that time. We therefore set $\Lambda =0$ in Eq.~(\ref{eha}).

 The exact form of the field $\phi$ depends on both the geometry and gauge fields that the scalar field couples to during the QCD phase transition. However, the process of bubble nucleation lasts for a very short period compared to the cosmological expansion rate. 
 Thus, the geometric effect can be neglected, and we will focus on the effects of the QCD nucleation only. Since QCD phase transition happens after inflation, the spacetime is approximately flat. Small perturbations around flat spacetime can be written as
\begin{equation}
g_{\mu\nu}=\eta_{\mu\nu}+h_{\mu\nu} .
\end{equation}
The perturbation, $h_{\mu\nu}$, can be written in the standard Hodge decomposition as
\begin{eqnarray}
&h_{tt}=-2{\cal A}\\
&h_{ti}=\mathfrak{B}_i^\perp+\triangledown_i {\cal B}\\
&h_{ij}=\mathcal{H}^\perp_{ij}+\triangledown_i\mathcal{E}_j^\perp+\triangledown_j\mathcal{E}_i^\perp+2\eta_{ij}{\cal C} +2(\triangledown_i\triangledown_j-\frac{1}{3}\triangledown^2){\cal D} .
\end{eqnarray}
The gauge invariant components are \cite{Mottola:2016mpl}
\begin{eqnarray}
&&\Upsilon_{\cal A}={\cal A}+\dot{{\cal B}}-\ddot{{\cal D}}\\
&&\Upsilon_{\cal C}={\cal C}-\frac{1}{3}\triangledown^2 {\cal D}\\
&&\psi_i^\perp=\mathfrak{B}^\perp_i-\dot{\mathcal{E}}_i^\perp\\
&&H_{ij}^\perp \rightarrow H_{ij}^\perp .
\end{eqnarray}
The first two scalar variables satisfy \cite{Mottola:2016mpl}
\begin{equation}
\Box \Upsilon_{\cal A} =\Box \Upsilon_{\cal C} =\frac{8\pi G b'}{3}\Box^2 \phi =0 ,
\end{equation}
which describes two kinds of the scalar gravitational waves in the flat space. Around the flat space, the equation of motion of $\phi$ is \cite{Mottola:2016mpl}
\begin{eqnarray}
\Box^2 \phi=\frac{1}{2}\Big(E-\frac{2}{3}\Box R +\frac{b}{b'}C^2+\frac{1}{b'}\sum_i \beta_i L_i\Big)=8\pi J .
\end{eqnarray}
Therefore, $\Upsilon_A$ and $\Upsilon_C$ are
\begin{equation}
\Upsilon_{\cal A}=\Upsilon_{\cal C} =-\frac{16\pi Gb'}{3}\int d^3 \mathbf{x} \frac{1}{|\mathbf{r}-\mathbf{x}|}J(\tilde{t},\mathbf{x}) .
\end{equation}
where  ${\tilde t}$ accounts for the time delay in propagation of the signal. The far field approximation gives
\begin{equation}
\label{wave}
\Upsilon_{\cal A}=\Upsilon_{\cal C} \approx -\frac{G}{3r}\int d^3 \mathbf{x}A_{anom} .
\end{equation}

In the effective QCD bag model with $\rho_{bag}=-p_{bag}=750MeV/fm^3$, $N_c=3$ and $N_f=2$, the value of the anomaly is  \cite{Mottola:2016mpl}
\begin{equation} \label{anomaly}
A_{anom} = (11 N_c - 2N_f ) \frac{\alpha_s}{24\pi}G_{\mu \nu}^a G^{a \mu \nu}  = (11 N_c - 2N_f ) \frac{\alpha_s}{24\pi} (\rho_{bag}-3p_{bag}) \approx   -4.8\times 10^{36} erg/cm^3 .
\end{equation}

The cause of the anomaly is that both the vector and axial currents are classically conserved for massless fermions, but the axial is not conserved at the quantum level. If fermions are massive, then axial current is not conserved even at the classical level. For massive fermions, one loop calculations indicate that the anomaly is suppressed by the fermion mass squared.  Since the quarks are not massless after the electroweak phase transition, we might have to take this suppression into account.
Most likely, the anomaly is still fully unsuppressed at the QCD phase transition, since the relevant quark masses are much smaller than the temperature at the QCD phase transition.
However, just to be on the safe side, we will introduce an optional cut-off in frequencies which preserves only the energy $\omega$ of the gravitational waves which is high enough so that the fermion mass can be neglected, i.e.
\begin{equation} \label{limit}
m_{u,d}\ll \omega ,
\end{equation}
where $m_{u,d}$ are $u$ and $d$ quarks masses  (i.e. in the standard model they are $2$MeV and $5$MeV respectively). At gravitational waves frequencies lower than $m_{u,d}$, the effect of anomaly might be suppressed by a factor of $(\omega/2m_{u,d})^2$.

In addition, we note that  there are models in which quarks are still massless during the QCD phase transition \cite{Iso:2017uuu}. In that case the suppression given by Eq.~(\ref{limit}) will not be present.

\section{Scalar gravitational waves from the QCD phase transition}

We are finally ready to estimate the parameters for the scalar gravitational waves produced during the QCD phase transition.
The QCD phase transition happens at the temperature $T_c\approx 150MeV$. This temperature is within the region of validity of the effective field theory that we used. The temperature today is $0.235meV$. Therefore the QCD phase transition happens at the redshift of $z\approx 6.3\times 10^{11}$. From Eq.~(\ref{wave}), the scalar gravitational wave amplitude from a single bubble is

\begin{equation}
\label{wave1}
\Upsilon_{\cal A}=\Upsilon_{\cal C} \approx \frac{G}{3rc^4}A_{anom}V_{bubble}=\frac{G}{3rc^4}A_{anom}V_0 f(t)
\end{equation}
where the time parameter, $t$, starts at the moment $t=t_i$ when the temperature of the universe is equal to the QCD phase transition temperature $T=T_{QCD}$. $V_{bubble}$ is given by Eqs.~(\ref{Vb}) and (\ref{Vb1}), depending on the period in question. The deflagration period will contribute more in the high frequency regime (smaller bubbles), but it turns out that the magnitude of the signal is too small to be observed, so we will proceed with Eq.~(\ref{Vb1}).  Therefore,  $V_0\approx \frac{4\pi}{3}(\frac{d_{nuc}}{2})^3$ is a single bubble's final volume before it collides with another hadron bubble and merges with it.
The anomaly $A_{anom}$ is given by Eq.~(\ref{anomaly}) with an overall  negative sign because the process of bubble nucleation removes free gluons from the space instead of creating them.  To introduce an optional cut-off in frequencies, we first perform a Fourier transform of the time domain function $f(t)$,
\begin{equation}
\hat{f}(\omega) =\int_\infty ^{-\infty}f(t)\exp(-i \omega t) dt .
\end{equation}
For more accurate results, one should consider the bubble's spatial distribution. But for a slowly expanding bubble, the spatial structure will not significantly affect the result. At $t=t_f$, the phase transition ends, and the space does not have free quarks and gluons, so $f(t_f)=1$. Therefore, we take only the time interval $t_i<t<t_f$.

As we explained at the end of section \ref{sgv}, we might need to cut-off the frequencies lower than quark masses, so the window function is
\begin{equation} \label{wf}
W(\omega) = \left\{
  \begin{array}{lr}
    \Big(\frac{\omega}{2MeV}\Big)^2 & , |\omega| < 2MeV\\
    1 & , 2MeV<|\omega|
  \end{array}
\right. .
\end{equation}
This function should be applied to $\hat{f}$ to remove the low energy modes as in Eq.~(\ref{limit}). However, this is not necessary if we believe that the anomaly is unsuppressed at QCD temperatures (which are much higher than the quark masses), and also in the models in which quarks are still massless during the QCD phase transition, so we will work both with and without it, i.e.

\begin{equation} \label{wf}
\bar{f}(t) = \left\{
  \begin{array}{lr}
    \frac{1}{2\pi}\int_{-\infty}^\infty\hat{f}W\exp(i \omega t) d\omega & \text{, if a window function is applied}\\
    \frac{1}{2\pi}\int_{-\infty}^\infty\hat{f}\exp(i \omega t) d\omega & \text{, if a window function is not applied}
  \end{array}
\right. ,
\end{equation}

The scalar gravitational wave amplitude  is now rewritten as
\begin{equation}
\Upsilon_A=\Upsilon_C \approx \frac{G}{3rc^4}A_{anom}V_0 \bar{f}(t)
\end{equation}

This is the gravitational amplitude from one single bubble. We will now include contribution from all of the bubbles, and the effect from the redshift. For  stochastic gravitational waves, the characteristic strain $h_c$ can be obtain from the power spectral density, $S_h$\cite{Moore:2014lga}, as
\begin{equation}
h_c=\sqrt{S_h \, \nu}.
\end{equation}
where $\nu$ is the gravitational wave frequency. Since $S_h$ is closely related to the energy density of gravitational waves, we will derive the energy density first and then find out the characteristic strain at the present time.

The energy momentum tensor for gravitational waves is
\begin{equation}
T_{\mu\nu}=\frac{c^4}{32\pi G}<\partial_\mu h_{\alpha\beta}\partial_\nu h^{\alpha\beta}>
\end{equation}
where the angle brackets denote averaging over several wavelengths.
The energy radiated by a single bubble can be estimated from energy flux, $T_{tr}$, as
\begin{eqnarray}
E_b &=& \frac{c^2}{32\pi G}\int \Upsilon_A \Upsilon_A k\omega dtdS\nonumber\\
&\approx &\frac{G}{72\pi c^5}A_{anom}^2V_0^2\int_0^\infty \hat{f}(\omega)\hat{f}^*(\omega)W^2(\omega)\omega^2d\omega ,
\end{eqnarray}
where $k$ is the wavenumber. For the integrated signal, we have to take into account all the bubbles, and also an appropriate energy redshift from the time of the signal creation till today.
The scalar gravitational waves energy density  at the time of the QCD phase transition was
\begin{equation}
\rho = nE_b ,
\end{equation}
where $n=d_{nuc}^{-3}$ is the bubble number density. Since the gravitons are massless particles, their energy density decreases as the universe expanding. At the present time the energy density in gravitational waves, $\rho_0$, is
\begin{equation}
\rho_0 = \frac{nE_b}{(1+z)^4} .
\end{equation}
The power spectral density is
\begin{eqnarray}
S_h(f)&=&\frac{4G}{\pi c^2}\frac{\delta\rho_0}{f^2\delta f}\nonumber\\
&=&\frac{4\pi n G^2}{9 c^7 }A_{anom}^2V_0^2\frac{\hat{f}(\omega)\hat{f}^*(\omega)W^2(\omega)}{1+z}
\end{eqnarray}
where we used $\nu =\frac{\omega}{2\pi (1+z)}$.
To obtain numerical values, we set $z\approx 6.3\times 10^{11}$, which is the redshift at the QCD phase transition, as explained below Eq.~(\ref{Vb}). As we noted before, to cover all the cases in the literature, we use two possible $d_{nuc}$ values, $1$cm and $1$m. The value for $A_{anom}$ is given in Eq.~(\ref{anomaly}).

 \begin{figure}[h!]
   \centering
\includegraphics[width=12cm]{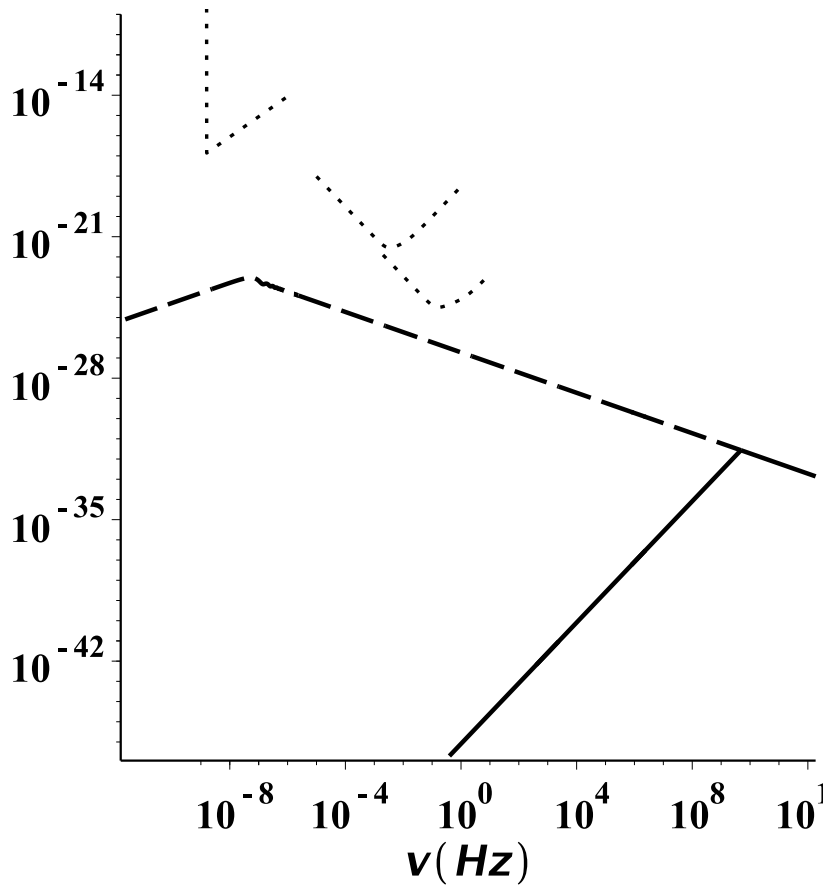}
\caption{The characteristic strain of the gravitational waves signal today, $h_c$, as a function of frequency, $\nu$. We set the value $d_{nuc} = 1$cm, which gives the smallest bubble volume and thus the weakest signal. The solid line is $h_c$ with the window function from Eq.~(\ref{wf}), while the dashed line is $h_c$ without this window function. The doted curves are the sensitivity regions of the detectors -- from low to high frequencies are SKA, LISA and BBO respectively.
}
\label{strain2}
\end{figure}

 \begin{figure}[h!]
   \centering
\includegraphics[width=12cm]{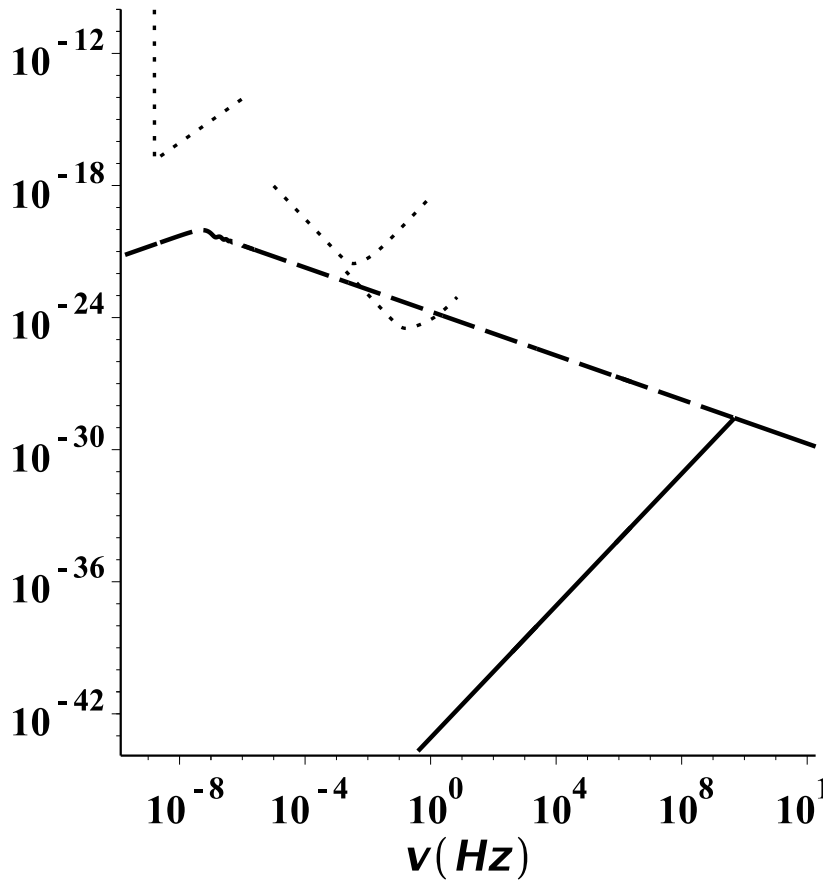}
\caption{The characteristic strain of the gravitational waves signal today, $h_c$, as a function of frequency, $\nu$. We set the value $d_{nuc} = 1$m, which gives larger bubble volumes and thus stronger signal. The solid line is $h_c$ with the window function from Eq.~(\ref{wf}), while the dashed line is $h_c$ without this window function. The doted curves are the sensitivity regions of the detectors -- from low to high frequencies are SKA, LISA and BBO respectively. Part of the signal is detectable by BBO. However, since the detector sensitivities are shown for the tensor modes, while it is known that LISA has an order of magnitude higher sensitivity to the scalar than to the tensor modes, the signal most likely falls within the LISA sensitivity region as well.}

\label{strain3}
\end{figure}

We plot the characteristic strain of the gravitational waves signal today, $h_c$, as a function of frequency, $\nu$, of scalar gravitational waves in Fig.~\ref{strain2} and \ref{strain3}. We give plots for two values of $d_{nuc}$, i.e. $1$cm and $1m$. The larger value of $d_{nuc}$ gives larger bubble volumes which in turn amplifies the anomaly effect, but reduces the bubble density. It turns out that the first effect is more important, so the the larger value of $d_{nuc}$ gives a stronger signal (Fig.~\ref{strain3}). It is notable that our signal is weaker than than the signal from the standard tensor modes \cite{Caprini:2007xq,Huber:2008hg,Jinno:2015doa}. This is because the tensor mode gravitational waves are created by very sudden change in bubbles energies and momenta during the collision. In contrast, the strength of the scalar mode depends on the phase transition rate rather than the rate of change of the matter energy and momentum. During the bubble's motion energy and momentum accumulate and get released at the moment of collision, however, one cannot accumulate the ``amount" of the QCD phase transition in a similar way. Eventually, motion of the bubble could increase the signal frequency via Doppler shift, but here we neglected this effect. One may also notice that the spectrum of the scalar mode decreases more slowly than for the usual transverse-tensor modes. This is because the QCD phase transition last longer than the bubble collision time scale, so it produces more low frequency modes.

We also show the case with the window function from Eq.~(\ref{wf}) which cuts off the frequencies lower than the quark masses, and also the case which includes all the frequencies (i.e. no window function).  The region of the parameter space with no window function is much more likely to be observed, especially if $d_{nuc}$ is large enough, since both the strain and the frequencies fall within the sensitivity of the near future gravitational wave experiments (e.g. The Big Bang Observer) (see e.g. Fig.~A1 in \cite{Moore:2014lga}). In addition, the detector sensitivities in fig. \ref{strain2} and \ref{strain3} are shown for the tensor modes. It is known that LISA has $10$ times higher sensitivity to the scalar mode than to the tensor modes\cite{Tinto:2010hz}. Thus, the signal most likely falls within the LISA sensitivity region as well.

\section{Conclusions}
In this paper we tried to connect the gravitational wave astronomy with fundamental particle physics.
The standard model of particle physics in the presence of gravity suffers from the well known trace anomaly. The origin of anomaly is purely quantum.
In the QCD sector, the anomaly gives rise to the new kind of (scalar) gravitational waves which are not present in the pure gravitational regime.
Quantum anomaly was originally derived for massless fermions, while the standard model quarks are massive. During the QCD phase transition, at temperatures higher than $150$ MeV, one can effectively neglect the u- and d-quark masses, and anomaly effects should become fully unsuppressed. Using the details of the first order phase transition, in particular the mechanism of the homogenous bubble nucleation, we were able to calculate the parameters relevant for the produced gravitational waves.
As the final result, we found the characteristic strain of the gravitational waves signal as it should look like today. To remain on the safe side, we introduced the window function which cuts-off very low  frequencies of the produced gravitational waves, where the anomaly calculations might not be completely trusted. For comparison, in Fig.~\ref{strain2} and \ref{strain3} we show  the characteristic strain both with and without the window function.  The region with no window function (i.e. no suppression in frequencies) is much more likely to be observed in near future gravitational wave experiments (e.g. LISA and The Big Bang Observer).  The interesting bottom line is that we could in principle learn something about the obscure quantum aspects of the standard model of particle physics using gravitational wave astronomy.


\begin{acknowledgments}
D.C Dai was supported by the National Science Foundation of China (Grant No. 11433001 and 11775140), National Basic Research Program of China (973 Program 2015CB857001) and  the Program of Shanghai Academic/Technology Research Leader under Grant No. 16XD1401600.
DS was partially supported by the US NSF grant PHY 1820738.
\end{acknowledgments}

\end{document}